\documentclass[%
 reprint,
 amsmath,amssymb,
 aps,
]{revtex4-2}

\usepackage{graphicx}
\usepackage{dcolumn}
\usepackage{bm}

\usepackage{braket}
\usepackage[dvipsnames]{xcolor}
\usepackage{soul}
\usepackage{natbib}

\begin{document}

\preprint{APS/123-QED}

\title{Recoil momentum of an atom absorbing light in a gaseous medium and the Abraham-Minkowski debate}

\author{João G. L. Condé}
 \email{joaoconde@ufmg.br}
\author{Pablo L. Saldanha}%
\affiliation{Departamento de Física, Universidade Federal de Minas Gerais, 30161-970 Belo Horizonte-MG, Brazil}

\date{\today}

\begin{abstract}
We discuss a fundamental question regarding the Abraham-Minkowski debate about the momentum of light in a medium: If an atom in a gas absorbs a photon, what is the momentum transferred to it? We consider a classical model for the internal degrees of freedom of the absorbing atom, computing the absorbed energy and momentum using the Lorentz force law due to the microscopic electromagnetic fields. Each non-absorbing atom from the gas is treated as a dielectric sphere, with the set of atoms forming a linear, dielectric, non-magnetic, and non-absorbing medium with a refractive index $n$ close to one. Our numerical results indicate that if the atoms are classically localized, the average absorbed momentum increases with $n$, but is smaller than Minkowski's momentum $np_0$, $p_0$ being the photon momentum in vacuum. However, experiments performed with Bose-Einstein condensates [Phys. Rev. Lett. $\mathbf{94}$, 170403 (2005)] are consistent with the atom absorbing Minkowski's momentum. We argue that there is no contradiction between these results since, in a Bose-Einstein condensate, the atoms are in a quantum state spatially superposed in a relatively large volume, forming a ``continuous'' medium. In this sense, the experimental verification of an atomic momentum recoil compatible with Minkowski's momentum would be a quantum signature of the medium state.

\end{abstract}

\maketitle

\section{Introduction}
The momentum of light in vacuum has been well-established since the end of the 19th century. However, when it comes to the momentum of light in material media, there has been a controversy for over one hundred years \cite{Brevik1979,Pfeifer2007,Barnett2010,Milonni2010,Kemp2011}. The two most common expressions are the ones formulated by Abraham and Minkowski \cite{Abraham1910,Minkowski1908}. They determine that the momentum density of light in a medium should be $\mathbf{E}\times\mathbf{H}/c^2$ and $\mathbf{D}\times\mathbf{B}$, respectively, where $\mathbf{E}$ is the electric field, $\mathbf{B}$ is the magnetic field, $\mathbf{D}=\epsilon_0\mathbf{E}+\mathbf{P}$, $\mathbf{H}=\mathbf{B}/\mu_0-\mathbf{M}$, $\mathbf{P}$ and $\mathbf{M}$ are the polarization and magnetization of
the medium, $\epsilon_0$ and $\mu_0$ are the permittivity and permeability
of free space, and $c$ is the speed of light in vacuum. An essential distinction between Abraham's and Minkowski's formulations is how the momentum of a light pulse depends on the refractive index $n$ of the medium. Minkowski's formulation of electrodynamics in continuous media predicts an electromagnetic momentum proportional to $n$, while Abraham's formulation predicts an electromagnetic momentum inversely proportional to $n$. This divergence originates what is now called the Abraham-Minkowski debate. 

Several experimental works in the past decades have been devoted to determining which expression is the correct description of the momentum density of light \cite{Jones1951,Ashkin1973,Walker1975,Walker1975-2,Jones1978,Gibson1980}. However, at the same time, it was understood that many different formulations for electrodynamics of continuous media, with different energy-momentum tensors and different expressions for the electromagnetic momentum density, lead to the same experimental predictions when the appropriate material energy-momentum tensor is taken into account \cite{Pfeifer2007,Penfield1967}. Recently, it has been argued that Abraham's expression is associated with the kinetic electromagnetic momentum density and Minkowski's with the canonical electromagnetic momentum density \cite{Barnett2010-2}. Other expressions also carry meaningful physical importance \cite{Pfeifer2007,Burt1973,Gordon1973,Loudon1997,Mansuripur2005,Loudon2004,Saldanha2010,Saldanha2011,Partanen2017,Anghinoni2022}, such as the momentum density $\epsilon_0\mathbf{E}\times\mathbf{B}$, which is compatible with momentum conservation in several situations when the material part of the momentum is computed employing the Lorentz force law \cite{Saldanha2010,Saldanha2011}. After more than 100 years, the Abraham-Minkowski debate is still a topic of interest, with theoretical \cite{Partanen2017,Anghinoni2022,Correa2016,Wang2016,Bliokh2017,Saldanha2017,Correa2020,Partanen2021,Partanen2023} and experimental \cite{Campbell2005,Astrath2014,Zhang2015,Schaberle2019,Astrath2022,Verma2023} works being published to clarify different aspects of the momentum of light in a medium. In particular, one work from Campbell \textit{et al.} \cite{Campbell2005} has shown that the momentum transferred from light to an atom from a Bose-Einstein condensate is proportional to $n$, as predicted by Minkowski's expression.

Here we aim to analyze the momentum recoil that light causes in an absorbing atom inside a linear, dielectric, non-magnetic, and non-absorbing gaseous medium, modeled by a set of rigid dielectric spheres, in the limit where the medium has a refractive index close to one. We consider a classical model for the internal degrees of freedom of the absorbing atom, computing the energy and momentum transfers by the Lorentz force law due to the action of the microscopic electromagnetic fields. We show that the average ratio between the momentum and the energy absorbed by the atom increases with $n$, but is smaller than $n/c$, such that the average momentum recoil is smaller than Minkowski's momentum. We argue that this result is not in conflict with the experiments from Campbell \textit{et al.} \cite{Campbell2005}, since in a Bose-Einstein condensate, the atoms are in a quantum superposition state in space, such that the model of rigid spheres is not valid and a continuous medium approximation, which predicts a ratio between the absorbed momentum and energy equal to $n/c$, is more reasonable. In this sense, the experiments from Campbell \textit{et al.} \cite{Campbell2005} can be seen as a signature of quantum properties of the atomic medium since we predict that experiments performed with a classical gas would result in a different average momentum transfer for the absorbing atom. 

\begin{figure}
    \centering
    \includegraphics[scale=0.375]{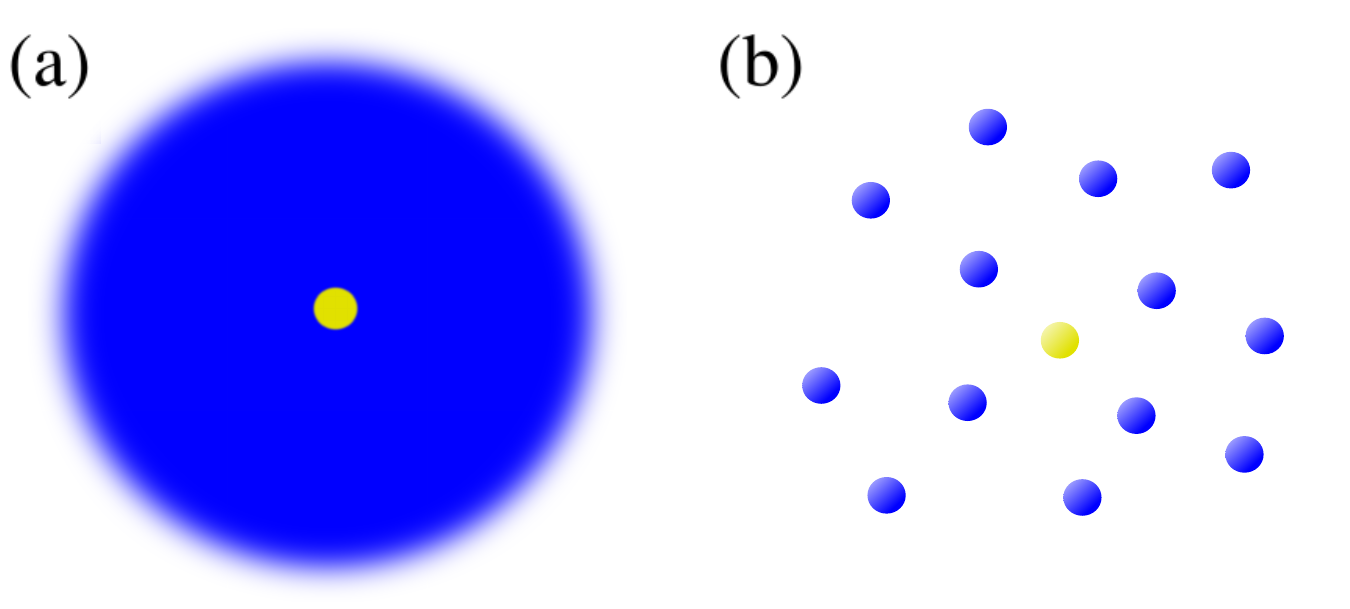}
    \caption{Different models for a gaseous medium where one atom absorbs light. The color
yellow indicates the atom that absorbs light, while the color blue
represents the material medium. (\textbf{a})  The absorbing atom
is immersed in a continuous medium. (\textbf{b})  The absorbing atom is immersed in a cloud of non-absorbing atoms modeled as rigid dielectric spheres.}
    \label{fig:1}
\end{figure}

\section{Microscopic model for a gaseous medium}

Fig. 1(a) depicts an atom (in yellow) immersed in a continuous, linear, dielectric, non-magnetic, and non-absorbing medium (in blue). The atom absorbs light, and we want to compute the ratio between the absorbed momentum and energy after this interaction. Fig. 1(b) shows a microscopic model where the medium is composed of linear, dielectric, non-magnetic, and non-absorbing atoms, modeled as spheres with radius $R$ (in blue), and the absorbing atom (in yellow). We show in the following sections that the microscopic model of Fig. 1(b) predicts different results compared to the continuous medium model of Fig. 1(a). In the present section, we discuss the microscopic model of Fig. 1(b) in detail.

When the non-absorbing atoms depicted in Fig. 1(b) are in a region containing an electric field, they polarize. The induced dipole moment in each atom $\mathbf{p}_i$ can be written as $\mathbf{p}_i(t)=\alpha\epsilon_0\boldsymbol{\mathcal{E}}(\mathbf{r}_i,t)$ using the usual complex notation, where $\boldsymbol{\mathcal{E}}(\mathbf{r}_i,t)$ is the microscopic value of the electric field at the position $\mathbf{r}_i$ of the center of the atom labeled by the index $i$ \cite{Fowles1989}. The parameter $\alpha$ is the atoms {polarizability}, and its magnitude is related to the strength of the interaction. In this model, the macroscopic polarization of the medium should come from the dipole moments of the atoms. The medium polarization $\mathbf{P}$ from $N$ dipoles with average dipole moment $\mathbf{p}_0$ in a volume $V$ is thus $\mathbf{P}=N\mathbf{p}_0/V$. Also, for linear dielectric media, the relation between the average electric field and polarization is $\mathbf{P}=\epsilon_0\chi\mathbf{E}$, where $\chi$ is the medium electric susceptibility, such that $\chi=N\alpha/V$. In this work, we consider that the medium refractive index is close to one, such that we may approximate 
\begin{align}
    n=\sqrt{1+\chi}\approx1+\chi/2=1+\frac{N\alpha}{2V}.
\end{align}
This result can also be obtained from the well-known Clausius-
Mossotti relation in the limit of refractive index close to one \cite{Jackson1999}. In this regime, the total electromagnetic fields produced by the atoms are considerably smaller than the incident fields, such that we may consider that only the incident electric field $\mathbf{E}_0$ induces polarization on the atoms, i.e.,  $\mathbf{p}_i(t)=\alpha\epsilon_0\mathbf{E}_0(\mathbf{r}_i,t)$. This approximation greatly simplifies the numerical calculations that will be presented.

\begin{figure}
    \centering
    \includegraphics[scale=0.298]{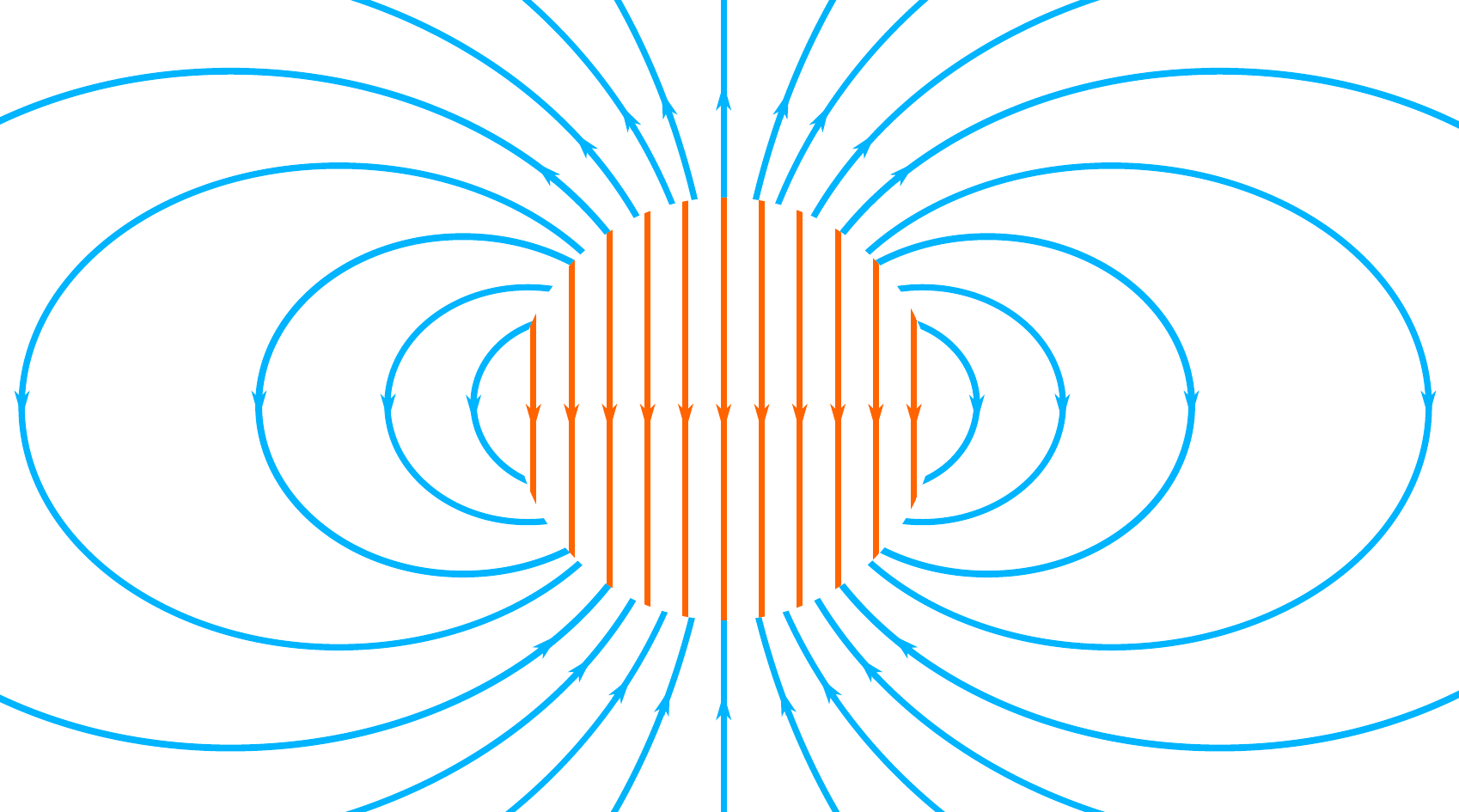}
    \caption{Electric field lines of a homogeneously polarized sphere in the static regime. The electric field inside the sphere (in orange) is homogeneous and in the opposite direction of the applied field $\mathbf{E}_0$ that induces the polarization. Outside the sphere, the electric field (in blue) is one of a perfect dipole at the sphere center.}
    \label{fig:dipolo}
\end{figure}

In the medium depicted in Fig. \ref{fig:1}(b), the total electromagnetic fields are equal to the superposition of the incident electric and magnetic fields $\mathbf{E}_0$ and $\mathbf{B}_0$ entering the medium and the electric and magnetic fields $\boldsymbol{\mathcal{E}}_i$ and $\boldsymbol{\mathcal{B}}_i$ generated by each atom. We consider the incident fields to be in the form
\begin{align}
    \mathbf{E_0}=E_0 e^{i(\mathbf{k}_0\cdot\mathbf{z}-\omega t)}\mathbf{\hat{x}},\;
    \mathbf{B_0}=B_0 e^{i(\mathbf{k}_0\cdot\mathbf{z}-\omega t)}\mathbf{\hat{y}},\label{2}
\end{align}
with $\mathbf{k}_0=(2\pi/\lambda_0)\mathbf{\hat{z}}$, $\lambda_0$ being the light wavelength in vacuum. In the present work, we model each non-absorbing atom as a linear, dielectric, non-magnetic, and non-absorbing sphere with a radius $R$. That is, to compute the fields that atom $i$ produces, we compute the fields that one of these dielectric spheres, placed in the same position and subjected to the same electromagnetic fields, would produce. Results shown ahead corroborate the validity of this model. The radius $R$ used for the spheres is much smaller than the wavelength of the external fields $\mathbf{E_0}$ and $\mathbf{B_0}$, such that the external fields are approximately homogeneous inside the atoms. When linear dielectric spheres are exposed to homogeneous oscillating electric fields, they respond producing electric and magnetic fields equivalent to an oscillating electric dipole in the region outside of the sphere \cite{Jackson1999}. So, if atom $i$ has a dipole moment given by $\mathbf{p}_i(t)=\alpha\epsilon_0\mathbf{E}_0(\mathbf{r}_i,t)$, we consider that, outside it, it produces electromagnetic fields
\begin{align}
    \boldsymbol{\mathcal{E}}_i=\frac{e^{ikr}}{4\pi\epsilon_0}\Big\{&\frac{\omega^2}{c^2r}(\hat{\mathbf{r}}\times\mathbf{p}_i)\times\hat{\mathbf{r}}+\nonumber\\&+\Big(\frac{1}{r^3}-\frac{i\omega}{cr^2}\Big)[3\hat{\mathbf{r}}(\hat{\mathbf{r}}\cdot\mathbf{p}_i)-\mathbf{p}_i]\Big\},\label{311}\\
     \boldsymbol{\mathcal{B}}_i=\frac{\omega^2e^{ikr}}{4\pi\epsilon_0 c^3}&\Big(\frac{1}{r}-\frac{c}{i\omega r^2}\Big)(\hat{\mathbf{r}}\times\mathbf{p}_i),\;\;\mathrm{for}\;r> R,\label{312}
\end{align}
where $\mathbf{r}$ is a distance vector from the center of the atom to the evaluated point \cite{Jackson1999}. Since the radius $R$ of the atom is considered to be much smaller than the light wavelength, for the region inside the atom, we disregard the radiating components of the field and consider that they produce a uniform electric field and a circulating magnetic field with zero average \cite{Jackson1999}. Since we are not interested in the spatial variation of the fields inside the atoms, we may consider a null magnetic field in our calculations. So, we consider that the field generated by each atom polarization inside it is
\begin{align}
     \boldsymbol{\mathcal{E}}_i=\frac{-\mathbf{p}_i}{4\pi\epsilon_0R^3},\;\;
     \boldsymbol{\mathcal{B}}_i=0,\;\;\mathrm{for}\;r< R,
\end{align}
as is the case for a dielectric sphere exposed to a uniform electric field \cite{griffiths}. Fig. (\ref{fig:dipolo}) illustrates the electrostatic contribution for the electric field generated by each atom, which is dominant for $r\ll\lambda_0$. The total electromagnetic fields in the medium can then be written as
\begin{align}
    \boldsymbol{\mathcal{E}}=\mathbf{E}_0+\sum_{i=1}^N  \boldsymbol{\mathcal{E}}_i,\;
    \boldsymbol{\mathcal{B}}=\mathbf{B}_0+\sum_{i=1}^N \boldsymbol{\mathcal{B}}_i.\label{tot_field1}
\end{align}

\begin{figure}
    \centering
    \includegraphics[scale=0.205]{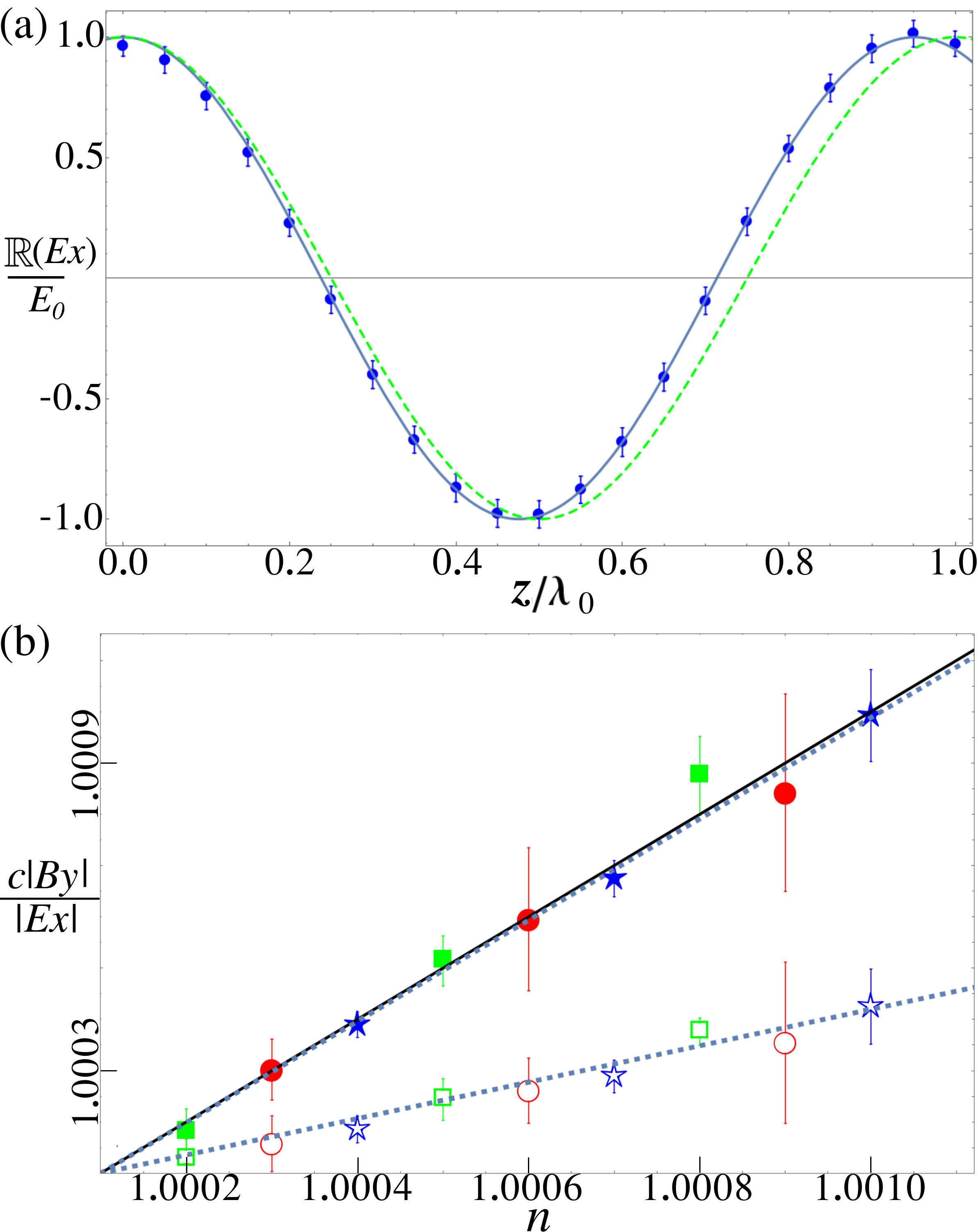}
    \caption{Numerical computations testing the model. (\textbf{a}) Real part of $E_x$ from Eq. (\ref{tot_field}) in a fixed time along the propagation direction ${z}$. The blue circles represent numerically calculated values for the field, and the solid blue line represents the fit curve $\cos[2\pi z/\lambda]$. The green dashed line represents the incident field $\mathbf{E}_0$ from Eq. (\ref{2}). The value $\lambda_0/\lambda=1.051\pm0.002$ obtained from the fit matches the expected value $\lambda_0/\lambda=n=1.050$ for the used parameters, which are $N=25625$, $\alpha=10^{-6}\lambda_0^3$, $V=25\lambda_0^3$, and $R=2\times10^{-2}\lambda_0$. (\textbf{b}) $|B_y|/|E_x|$, where $E_x$ and $B_y$ are computed from Eq. (\ref{tot_field}), for several values of the index of refraction $n$. The markers represent the numerically calculated values, and the solid black line is the theoretical prediction for continuous media $|B_y|/|E_x|=n/c$. The hollow markers present data filtered only to include positions outside all atoms, while the filled markers present unfiltered data. The dashed lines are the fits for the expression $c|B_y|/|E_x|=a(n-1)+1$, and the values obtained for $a$ are $a=0.99\pm0.02$ for unfiltered data and $a=0.36\pm0.01$ for the filtered data. The parameters used were $\alpha=10^{-6}\lambda_0^3$ and $R=2\times10^{-2}\lambda_0$ in green squares; $\alpha=2\times10^{-6}\lambda_0^3$ and $R=10^{-2}\lambda_0$ in red circles; $\alpha=5\times10^{-7}\lambda_0^3$ and $R=3\times10^{-2}\lambda_0$ in blue stars. We have $V=25\lambda_0^3$ in all cases, and the value of $n$ is varied through a variation of the number of atoms $N$. The error bars were computed by repeating the process five times with different drawn positions and computing the standard deviation.
    }
    \label{fig:3}
\end{figure}

We expect the averaged macroscopic fields to be
\begin{align}
    \mathbf{E}=\Big<\mathbf{E}_0+\sum_{i=1}^N  \boldsymbol{\mathcal{E}}_i\Big>,\;\;
    \mathbf{B}=\Big<\mathbf{B}_0+\sum_{i=1}^N \boldsymbol{\mathcal{B}}_i\Big>.\label{tot_field}
\end{align}
The brackets represent the average over a region containing many atoms in a volume $\Delta V$ much smaller than the light wavelength cubed. We perform a numerical computation of the above equation to verify this relation between the microscopic and macroscopic fields. Using the software Wolfram Mathematica \cite{Mathematica}, we randomly draw the positions on which the $N$ atoms will be placed. Those positions are drawn in a volume in the form of a rectangular parallelepiped with size $\lambda_0$ in the $z$ direction and a square cross-section with sides $5\lambda_0$ in the $xy$ plane. The atomic densities we use in our numerical computations are between the one from the air at atmospheric pressure and room temperature (around $10^{19}$ molecules per cm$^3$) and the ones obtained in Bose-Einstein condensates (around $10^{13}$ atoms per cm$^3$).
The average electric and magnetic fields are computed by calculating the fields of Eq. (\ref{tot_field}) in $10^6$ random positions inside a small volume $\Delta V$. Note that some random positions will fall inside one atom, while others will fall outside all. Some results are shown in Fig. \ref{fig:3}. In Fig. \ref{fig:3}(a), we see that the wavelength of the average electric field respects the relation $\lambda\approx\lambda_0/n$. The filled markers in Fig. \ref{fig:3}(b) show that the ratio between the average magnetic and electric fields agrees with the expected relation between these fields $|B_y|/|E_x|=n/c$. These results corroborate the validity of the model. It is important to stress the contribution of the electric field generated by the atom polarization inside it, which points in the opposite direction of $\mathbf{E}_0$ as seen in Fig. (\ref{fig:dipolo}) and contributes for lowering the average electric field. As shown by the hollow markers in Fig. \ref{fig:3}(b), if only points outside the atoms are considered for computing the average fields, the ratio $|B_y|/|E_x|$ becomes smaller than $n/c$ and the continuous medium result is not recovered. 

\section{Momentum and energy transferred to the absorbing atom}\label{sec2}

In this section, we compute the amount of momentum and energy transferred by the electromagnetic fields to an atom absorbing light inside a gaseous medium described by the model presented in the previous section. We use the Lorentz force law
\begin{align}\label{lorentz}
    \mathbf{F}&=\frac{1}{2}\mathbb{R}\Big\{(\mathbf{p}^{*}\cdot\nabla) \boldsymbol{\mathcal{E}}+\frac{\partial\mathbf{p}^{*}}{\partial t}\times\boldsymbol{\mathcal{B}}\Big\},
\end{align}
where $\mathbf{p}$ is the dipole moment of the absorbing atom and $\mathbb{R}(x)$ denotes the real part of $x$, to compute the momentum transfer. $ \boldsymbol{\mathcal{E}}$ and $ \boldsymbol{\mathcal{B}}$ are the microscopic electric and magnetic fields given by Eq. (\ref{tot_field1}) at the atom position. The above force represents the average force in one optical cycle. We consider that the incident light is resonant with the atom. In this way, the dipole moment of the absorbing atom is related to the microscopic electric field as
\begin{align}\label{3}
    \mathbf{p}=i\alpha_0\epsilon_0 \boldsymbol{\mathcal{E}}.
\end{align}
Note that there is a phase difference of $\pi/2$ between the dipole oscillation and the electric field oscillation, which results in the atom's permanent absorption of momentum and energy. Note also that the atomic polarizability $\alpha_0$ differs from the polarizability $\alpha$ of the medium's atoms. Substituting Eq. (\ref{3}) into Eq. (\ref{lorentz}), we arrive at the following expression for the momentum transferred to the atom:
\begin{align}\label{4}
    \boldsymbol{\mathcal{P}}=\frac{\alpha_0\epsilon_0}{2}\int dt\mathbb{R}\left\{-i( \boldsymbol{\mathcal{E}}^{*}\cdot\nabla) \boldsymbol{\mathcal{E}}+\omega \boldsymbol{\mathcal{E}}^{*}\times\boldsymbol{\mathcal{B}}\right\}.
\end{align}

The absorbed energy can be computed through the work done by the electric field into the electric current associated with the time variation of the atomic dipole moment:
\begin{align}\label{energy}
    U=\frac{1}{2}\int dt \mathbb{R}\left\{\boldsymbol{\mathcal{E}}^{*}\cdot\frac{\partial\mathbf{p}}{\partial t}\right\}=\frac{\alpha_0\epsilon_0\omega}{2}\int dt \boldsymbol{\mathcal{E}}^{*}\cdot \boldsymbol{\mathcal{E}}.
\end{align}
If we set $U=\hbar\omega$, $\hbar\omega$ being the energy of one photon, the modulus of the momentum transfer according to Minkowski's expression would be $n\hbar\omega/c$, such that $\mathcal{P}/U=n/c$.

This calculation for the microscopic model is fundamental, coming only from the microscopic Lorentz force law. It does depend on neither Abraham's nor Minkowski's formulation since they are fundamentally macroscopic.

\begin{figure}
    \centering
    \includegraphics[scale=0.21]{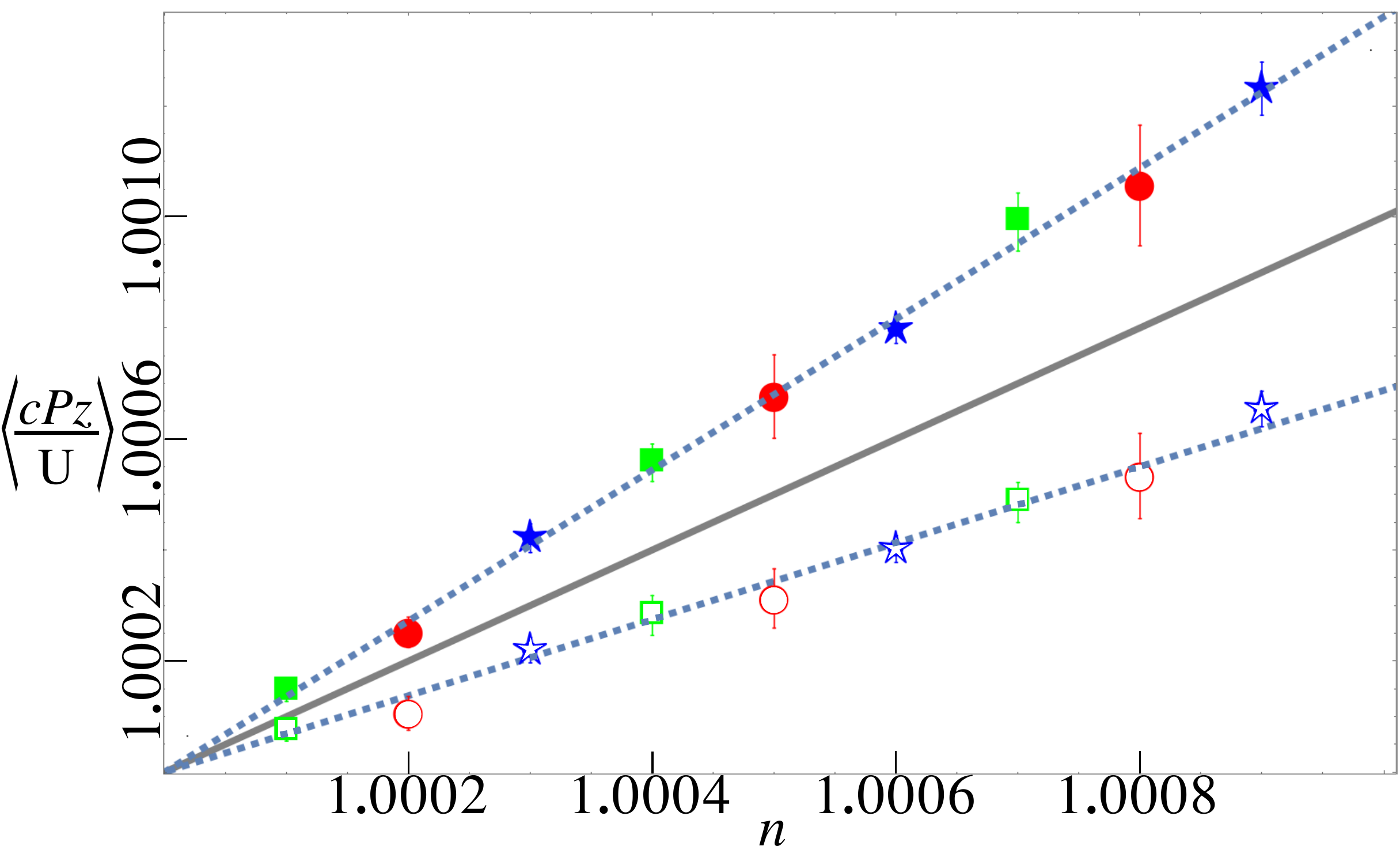}
    \caption{Average ratio between the $z$ component of the absorbed momentum $\mathcal{P}_z$ from Eq. (\ref{4}) and the absorbed energy $U$ from Eq. (\ref{energy}) as a function of $n$. The continuous black line is the function $\langle c\mathcal{P}_z/U\rangle=n$. The filled markers are acquired by considering positions for the absorbing atom inside and outside the non-absorbing atoms that compose the medium, and the hollow markers are acquired by filtering only positions that fall outside the atoms. The dotted lines are linear fits for each scenario. The expression used was $\langle c\mathcal{P}_z/U\rangle=a(n-1)+1$, and the values obtained for $a$ are $a=1.36\pm0.01$ for the unfiltered data and $a=0.69\pm0.01$ for the filtered data. Sets of points of each color and shape represent data acquired using different parameters in the computation, showing the independence of the results on the specific parameters used. The volume $V=25\lambda_0^3$ is the same for all sets. We have $\alpha=10^{-6}\lambda_0^3$ and $R=2\times10^{-2}\lambda_0$ in green squares; $\alpha=2\times10^{-6}\lambda_0^3$ and $R=10^{-2}\lambda_0$ in red circles; $\alpha=5\times10^{-7}\lambda_0^3$ and $R=3\times10^{-2}\lambda_0$ in blue stars. The index of refraction is varied by changing the number of atoms $N$. The bigger error bar for the data in red circles is associated with the smaller number $N$ of atoms in the medium for the particular set of parameters. In each run, $10^6$ random positions for the absorbing atom are chosen, and the error bars were computed by repeating the process five times with different drawn positions and computing the standard deviation.}
    \label{fig:4}
\end{figure}

The microscopic fields from Eq. (\ref{tot_field1}) are computed numerically using the same parameters as before. The momentum and energy acquired by the absorbing atom in a particular position are computed by Eqs. (\ref{4}) and (\ref{energy}). Different positions for the absorbing atom are randomly chosen, so the average ratio between the absorbed momentum $\boldsymbol{\mathcal{P}}$ and the absorbed energy $U$ can be computed for different medium parameters. We separately analyze the data of the situations where the observed point is outside all the atoms. This analysis is done considering that our model for the gas consists of a set of rigid dielectric spheres. This condition is reasonable for describing a classical gas of atoms, since the absorbing atom cannot be inside another atom in this case. Fig. \ref{fig:4} presents some results as hollow markers. We see that the computed average ratio $\langle\mathcal{P}_z/U\rangle$ is different from the one obtained with Minkowski's expression for the momentum transfer, having a smaller value. One could wonder if considering contributions from positions inside the atoms would result in Minkowski's momentum, but this is not the case, as shown by the filled markers in Fig. \ref{fig:4}. In this case, the average ratio $\langle\mathcal{P}_z/U\rangle$ becomes larger than $n/c$.

\begin{figure}
     \centering
     \includegraphics[scale=0.2015]{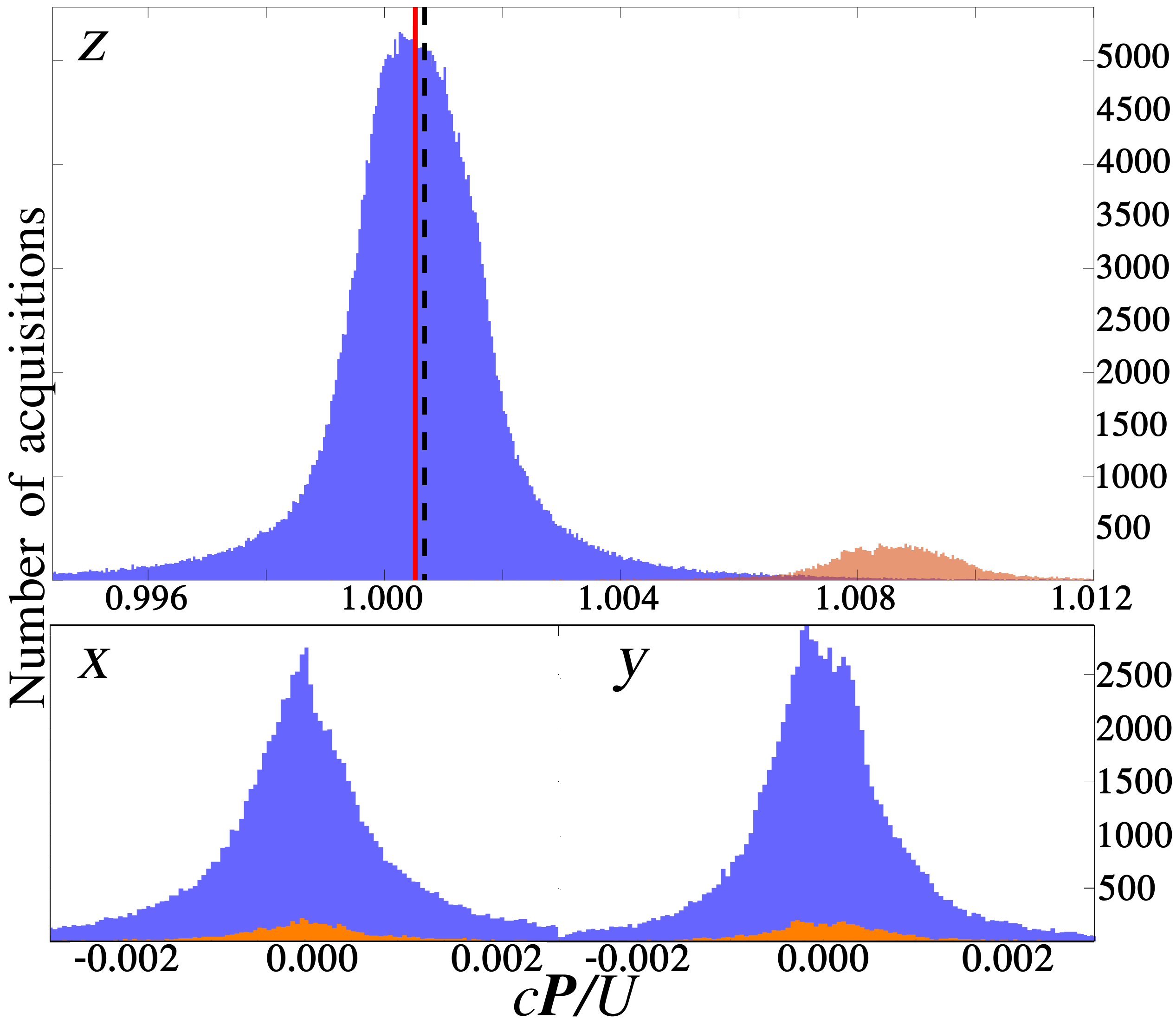}
     \caption{Number of each obtained value for the components of $c\boldsymbol{\mathcal{P}}/U$ for $n=1.00070$. The blue bars are related to positions for the absorbing atom that fall \textit{outside} all the atoms. The orange bars are related to positions for the absorbing atom that fall \textit{inside} one atom. The red vertical line (solid) in the first panel corresponds to the mean value of the distribution of the blue bars ($1.00048\pm0.00004$). In contrast, the black vertical line (dashed) corresponds to the prediction using Minkowski's momentum $c\mathcal{P}_z/U=n=1.00070$. The used parameters are $\alpha=2\times10^{-6}\lambda_0^3$, $V=25\lambda_0^3$, $R=2\times10^{-2}\lambda_0$ and $N=20008$.}
     \label{fig:5}
 \end{figure}

Fig. \ref{fig:5} shows the number of obtained values for the components of the ratio between the absorbed momentum $\boldsymbol{\mathcal{P}}$ and absorbed energy $U$ for one particular set of parameters when the position of the absorbing atom is randomly chosen in $10^6$ different positions. The distributions have relatively large variances since the microscopic fields have significant variations in the microscopic scale. It is valid to stress that since the $x$ and $y$ components of $\boldsymbol{\mathcal{P}}$ are much smaller than the $z$ component, they have a negligible influence on its modulus. So, graphics like the ones from Fig. \ref{fig:4} for  $\langle c|\boldsymbol{\mathcal{P}}|/U\rangle$ would be visually identical to the ones shown.

\section{Discussion}

Experiments performed with Bose-Einstein condensates show results consistent with the atoms acquiring Minkowski's momentum through light scattering \cite{Campbell2005}. Since light scattering can be seen as a photon absorption followed by one photon emission, this result differs from the one we present. Despite this fact, we argue that there is no contradiction between these results. In a Bose-Einstein condensate, the atoms are in a quantum state spatially superposed in a relatively large volume, forming a ``continuous'' medium. In a continuous medium, the fields are homogeneous, such that the term $(\boldsymbol{\mathcal{E}}^{*}\cdot\nabla) \boldsymbol{\mathcal{E}}$ in Eq. (\ref{4}) does not contribute for the momentum absorbed by the atom, contrary to the case in our calculations. Also, the microscopic and macroscopic fields are always equal without microscopic variations. That leads us to the simplified relation
\begin{align}\label{6}
    \frac{\boldsymbol{\mathcal{P}}}{U}&=\frac{\int dt\mathbb{R}\{\mathbf{E}^{*}\times\mathbf{B}\}}{\int dt\mathbf{E}^{*}\cdot\mathbf{E}}=\frac{B}{E}\hat{\mathbf{k}}=\frac{n}{c}\hat{\mathbf{k}}.
\end{align}
Note also that the ratio $\mathcal{P}_z/U$ is independent of the microscopic position of the absorbing atom in this continuous medium case. So, the distribution shown in Fig. \ref{fig:5} would have a sharp peak around $c\mathcal{P}_z/U=n$ instead of the large distribution that happens in the case of a classical gas of atoms. This sharp distribution for $c\mathcal{P}_z/U$ is consistent with the experimental results with Bose-Einstein condensates \cite{Campbell2005}.

It is also important to mention that, in this work, we have focused on the energy and momentum transferred by light to the one absorbing atom. Considering the superposition of the incident field with the field generated by the absorbing atom in the atoms nearby, there may be a permanent momentum transfer to these other atoms, even if they do not absorb light. The investigation of these permanent momentum transfers to the non-absorbing atoms, as well as a complete momentum balance analysis in the process, is outside the scope of the present work and should be analyzed in future works. Our microscopic model, in principle, could also be extended to treat other situations relevant in the Abraham-Minkowski debate, such as the momentum balance of light and matter at the interface between two different media \cite{Ashkin1973,Gordon1973,Astrath2014,Wang2016,Schaberle2019,Chesneau2022,Astrath2022,Verma2023}, the momentum balance in the reflection of light by a mirror immersed in a dielectric medium \cite{Jones1951,Jones1978,Correa2016,Gordon1973}, and the optical electro- and magnetostriction effects \cite{Hakim1962,Wang2016,Astrath2014,Astrath2022,Partanen2023}.

To conclude, we have considered the situation where an atom absorbs light immersed in a linear, dielectric,  non-magnetic, and non-absorbing gaseous medium modeled as a set of rigid dielectric spheres. We have computed the momentum and energy absorbed by the atom in numerical computations, which show that the average ratio between the absorbed momentum and energy is smaller than the one predicted using Minkowski's momentum. We thus predict that an experiment similar to the one from Campbell \textit{et al.} \cite{Campbell2005}, if performed with a classical gas instead of a Bose-Einstein condensate, would present different results for the atomic momentum recoil. In this sense, the experimental verification of an atomic momentum transfer proportional to $n$ shown in Ref. \cite{Campbell2005} can be considered a quantum signature of the medium state.

\begin{acknowledgments}
This work was supported by the Brazilian agencies CNPq (Conselho Nacional de Desenvolvimento Cient\'ifico e Tecnol\'ogico), CAPES (Coordenação de Aperfeiçoamento de Pessoal de Nível Superior), and FAPEMIG (Fundação de Amparo à Pesquisa do Estado de Minas Gerais).
\end{acknowledgments}


%

\end{document}